\documentclass{article}

\usepackage{graphicx}
\usepackage{amsmath}
\usepackage{amssymb}
\usepackage{url}
\usepackage{wasysym}
\usepackage{longtable}

\newtheorem{theorem}{Theorem}
\newtheorem{definition}{Definition}

\newtheorem{claim}[theorem]{Claim}

\newenvironment*{proof}[1][]{{\em Proof{#1}.}\ \ }{\hfill$\Box$}

\newcommand{\bbrd}[1]{\mbox{\rm{I}\kern-.1667em{#1}}}
\newcommand{\EXP}{\mathbb{E}}
\newcommand{\PROB}{\mathbb{P}}

\newcommand{\offather}{\mathrm{A}}
\newcommand{\ofmother}{\mathrm{B}}

\newcommand{\qsnp}{\mu_2}

\newcommand{\npsnp}{\bar{\mu}}
\newcommand{\nqsnp}{\bar{\mu}_2}
\newcommand{\ntsnp}{\bar{\mu}_3}
\newcommand{\nfsnp}{\bar{\mu}_4}
\newcommand{\ndeux}{\bar{\nu}_2}
\newcommand{\ntrois}{\bar{\nu}_3}
\newcommand{\cdeux}{\nu_2}
\newcommand{\ctrois}{\nu_3}

\newcommand{\mfreq}{\textbf{F}}

\newcommand{\mafcnt}{\omega}

\newcommand{\smpstat}{\tau}
\newcommand{\avgfreq}{\bar{f}}
\newcommand{\convprob}{\xrightarrow{\mathrm{p}}}

\begin{document}

\title{Non-identifiability of identity coefficients 
	at biallelic loci}
\author{Mikl\'os Cs\H{u}r\"os\thanks{Department of Computer Science and Operations Research, University of Montr\'eal; 
E-mail: csuros@iro.umontreal.ca.}}

\maketitle 

\begin{abstract}
	Shared genealogies introduce allele dependencies in diploid
	genotypes, as alleles within an individual or between
	different individuals will likely match when they originate
	from a recent common ancestor. 
	At a locus shared by a pair of diploid individuals, there are nine 
	combinatorially distinct modes of identity-by-descent (IBD), 
	capturing all possible combinations of coancestry and inbreeding. 
	A distribution over the IBD modes is described by the nine associated 
	probabilities, known as (Jacquard's)  identity coefficients.
	The genetic relatedness between two 
	individuals can be succinctly characterized by the identity coefficients 
	corresponding to the joint genealogy. 
	The identity coefficients (together with allele frequencies) 
 	determine 
 	the distribution of joint genotypes at a locus. 
	At a locus with
	two possible alleles, identity coefficients 
	are not identifiable 
	because different coefficients can generate the same
	genotype distribution.
	
	We analyze precisely how different IBD modes combine
	into identical genotype distributions at diallelic loci. 
	In particular, we describe IBD mode mixtures that 
	result in identical genotype distributions at 
	all allele frequencies, implying the non-identifiability  
	of the identity coefficients from independent loci.  
	Our analysis yields an exhaustive characterization of
	relatedness statistics that are always identifiable. 
	Importantly, we show that identifiable relatedness
	statistics include the kinship coefficient (probability that
	a random pair of alleles are identical by descent between
	individuals) and inbreeding-related
	measures, which can thus be estimated from genotype
	distributions at independent loci.
\end{abstract}

\section{Introduction}
Non-random mating histories, selection, finite population sizes, and many other  
causes create dependencies between alleles in a diploid population. 
Because of joint genealogies, alleles may match 
within or across genotypes for being unmodified copies of a common ancestral state. 
Such alleles are said to be {\em identical by descent} \cite{Malecot,Jacquard}. 
Combinatorially distinct partitionings of identical-by-descent (IBD) alleles 
for a pair of unordered diploid genotypes 
are called {\em identity modes}~\cite{Jacquard}. 
The two individuals' joint pedigree 
defines the possible inheritance histories for the four alleles,
which combine into a distribution over the identity modes.  
Every identity mode generates its own probability distribution over 
the joint genotypes at a locus, and the observable genotypic distribution is 
the mixture of the mode-specific distributions. 
The probabilities of the identity modes, or {\em identity coefficients},
characterize thus the individuals' genetic relatedness succinctly.

\begin{figure}
\centerline{\includegraphics[width=\textwidth]{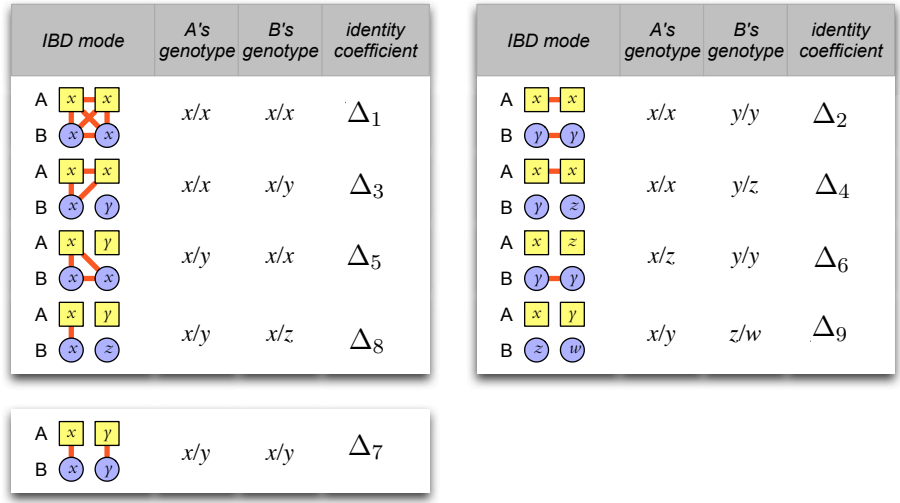}}
\caption{\textbf{Identity modes for a diploid genotype pair}. Thick red lines mark identity by descent. 
Alleles $x,y,\dotsc$ observed in the genotypes may be equal. Probabilities for 
different modes are denoted by the identity coefficients $\Delta_i$. 
} \label{fig:coancestry}
\end{figure}

The identity coefficients can be computed for any known pairwise genealogy \cite{Cockerham.ibd.computation,Lange}. 
Hypothetical pedigrees can be thus assessed by 
comparing implied genotype distributions with empirical ones \cite{Thompson.pairwise,Milligan.relatedness.ML}. 
But can identity coefficients be directly inferred from genotype distributions without genealogies? 
The answer depends on the number of alleles. There are nine identity modes 
for a pair of diploid individuals (Figure~\ref{fig:coancestry}), which define eight independent 
identity coefficients; the ninth one is implied since the coefficients sum to one.  
At loci with only two alleles, nine genotype pairs are possible, but because of redundancy, there are not 
enough many different genotype pairs to make certain inference possible: more than one set of coefficients generate 
the same joint genotype distribution. 
Genotype distributions at loci with three or more alleles,
however, convey enough information in principle to identify a 
single set of identity coefficients that produce it.
Among molecular markers, multiallelic microsatellite loci provide in consequence 
high discriminatory power for a detailed characterization of 
genetic relatedness, but diallelic single-nucleotide and insertion-deletion 
have restricted utility \cite{Weir.relatedness.markers}. 
Here, we scrutinize the inherent ambiguity of relatedness in diallelic genotypes. 
Specifically, our aim is to find what aspects of coancestry result in non-identifiability and 
to characterize statistical measures of the identity mode distribution that can be 
consistently estimated from joint genotype frequencies. 

\section{Theory}
\subsection{Identity coefficients and biallelic genotype distributions}\label{ss:ibd}
{\em Identity by descent} \cite{Malecot} encapsulates the dependence between 
diploid genotypes due to shared parentage. 
Two alleles are identical by descent (IBD) if they originate from a common ancestral allele without modification. 
Equivalence relations for four alleles of two diploid genotypes take one of nine combinatorially distinct 
forms~\cite{Harris.inbred.relatives,Jacquard,Lange}, or {\em identity modes},
as illustrated in Figure~\ref{fig:coancestry}.

The individuals' joint pedigree determines the possible identity modes and their associated frequencies, 
specified by the vector of coefficients $\Delta_i\colon i=1,\dotsc,9$ 
using the notation of Jacquard~\cite{Jacquard}. For instance, 
children of the same parents from non-overlapping lineages
inherit two IBD alleles with probability $\Delta_7=\tfrac{1}{4}$, 
one IBD set from either parent with probability $\Delta_8=\tfrac{1}{2}$, 
and four independent alleles with probability $\Delta_9=\tfrac{1}{4}$. 

\begin{table}
\caption{\textbf{Distribution of biallelic genotypes by identity mode}}\label{tbl:genotype.twoalleles}
\footnotesize
\begin{tabular}{lccccccccc}
Mode&0/0:0/0&1/1:1/1&1/1:0/1&0/1:1/1&0/1:0/1&1/1:0/0&0/0:1/1&0/1:0/0&0/0:0/1 \\
\hline
1	& $q$	& $p$	& 0 	& 0 	& 0 	& 0 	& 0 	& 0 	& 0 \\
2	& $q^2$	& $p^2$ & 0 	& 0 	& 0 	& $pq$	& $pq$	& 0 	& 0 \\
3	& $q^2$	& $p^2$	& $pq$	& 0 	& 0 	& 0 	& 0 	& 0 	& $pq$\\
4	& $q^3$	& $p^3$	&$2p^2q$& 0	 	& 0 	& $pq^2$& $p^2q$& 0 	& $2pq^2$\\
5	& $q^2$ & $p^2$ & 0 	& $pq$	& 0 	& 0 	& 0 	& $pq$ 	& 0 \\
6	& $q^3$ & $p^3$ & 0 	&$2p^2q$& 0 	& $p^2q$& $pq^2$&$2pq^2$& 0 \\
7	& $q^2$ & $p^2$ & 0 	& 0 	& $2pq$	& 0 	& 0 	& 0 	& 0 \\
8 	& $q^3$ & $p^3$ & $p^2q$& $p^2q$& $pq(p+q)$	& 0	& 0 	& $pq^2$ & $pq^2$\\
9 	& $q^4$ & $p^4$ &$2p^3q$&$2p^3q$&$4p^2q^2$ 	&$p^2q^2$	& $p^2q^2$	& $2pq^3$	& $2pq^3$ \\
\hline
\end{tabular}

\begin{flushleft}
Genotypic probabilities for every mode are given by assuming that 
alleles are chosen independently for each IBD group, with probability~$p$ for allele~1 (minor allele) 
and with probability~$q$ for allele~0 (major allele). Genotypes are unordered ($1/0$ and $0/1$ 
are considered equivalent). 
\end{flushleft}
\end{table}

Suppose that the locus has two alleles, and alleles~1 and~0 (minor and major) occur with frequencies~$p$ and~$q$, respectively
in both individuals.
Every mode generates its own conditional distribution of joint genotypes. In mode~8, the individuals are~$0/1$ 
heterozygotes simultaneously with 
probability $pq^2+p^2q$ since either the IBD alleles are the mutants, or two mutant alleles are chosen independently.  
In contrast, if all four alleles are sampled independently (identity mode~9) then 
the joint genotype $0/1:0/1$ occurs with probability~$4p^2q^2$, accounting for two minor and two major 
alleles in~4 possible orderings. Table~\ref{tbl:genotype.twoalleles} lists the 
complete set of genotypic probabilities. 
Denote the distribution of joint genotypes by 
\[
\mathbf{f}=(f_{0000}, f_{1111}, f_{1101}, f_{0111}, f_{0101}, f_{1100}, f_{0011}, f_{0100}, f_{0001}).
\]

Table~\ref{tbl:genotype.twoalleles} corresponds to the system of equations 
\begin{equation}\label{eq:jointcalls.p}
\begin{aligned}
f_{0000} & = q\Delta_1 + q^2\bigl(\Delta_2+\Delta_3+\Delta_5 +\Delta_7\bigr)
	+q^3\bigl(\Delta_8+\Delta_4+\Delta_6\bigr)+q^4\Delta_9 \\
f_{1111} & = p\Delta_1 + p^2\bigl(\Delta_2+\Delta_3+\Delta_5 +\Delta_7\bigr)
	+p^3\bigl(\Delta_4+\Delta_6+\Delta_8\bigr)+p^4\Delta_9\\
f_{1101} & = pq\bigl(\Delta_3 + p(\Delta_8+2\Delta_4+2p\Delta_9)\bigr) \\
f_{0111} & = pq\bigl(\Delta_5 + p(\Delta_8+2\Delta_6+2p\Delta_9)\bigr) \\
f_{0101} & = pq\bigl(2\Delta_7+ q\Delta_8+4pq\Delta_9\bigr) \\
f_{1100} & = pq\bigl(\Delta_2 + q\Delta_4+p\Delta_6+pq\Delta_9\bigr) \\
f_{0011} & = pq\bigl(\Delta_2 + q\Delta_6+p\Delta_4+pq\Delta_9\bigr) \\
f_{0100} & = pq\bigl(\Delta_5 + q\Delta_8 + 2q\Delta_6 + 2q^2\Delta_9\bigr) \\
f_{0001} & = pq\bigl(\Delta_3 + q\Delta_8 + 2q\Delta_4 + 2q^2\Delta_9\bigr). 
\end{aligned}
\end{equation}
In matrix form,
\begin{equation}\label{eq:linear.twoalleles}
\mathbf{f} = \mfreq \cdot \boldsymbol{\Delta}, 
\end{equation}
and 
Table~\ref{tbl:genotype.twoalleles} gives the transpose of~$\mfreq$. 
Note that the matrix structure guarantees $\sum_i \Delta_i=1$ when 
$f_{0000}+f_{1111}+\dotsc+f_{0001}=1$ since the 
all-1 row vector $\mathbf{e}=\begin{pmatrix} 1 & 1 & \dotsm & 1 \end{pmatrix}$ is a left eigenvector: 
\[
1 = \mathbf{e}\cdot \mathbf{f} = \mathbf{e} \cdot \mfreq \cdot \boldsymbol{\Delta}  = \mathbf{e} \cdot \boldsymbol{\Delta}.
\]

The matrix~$\mfreq$ projects the vector of identity coefficients~$\boldsymbol{\Delta}$ to the 
vector of genotype probabilities~$\mathbf{f}$. Consequently,  
identity coefficients can be inferred from the biallelic genotype distribution 
if and only if the matrix~$\mfreq$ is invertible. 
The matrix rows are, however, linearly dependent. 

\begin{claim}\label{cm:mafcnt}
When $p+q=1$, dependencies between genotype probabilities include the following two. 
\begin{gather}
f_{1101}+2f_{1100}+f_{0100}  = f_{0111}+2f_{0011}+f_{0001};\label{eq:mutantdiff}\\
p = \begin{aligned}[t] 
	& f_{1111} + \frac34 \bigl(f_{1101}+f_{0111}\bigr)\\
	+& \frac12 \bigl(f_{0101}+f_{1100}+f_{0011}\bigr) + \frac14\bigl(f_{0100}+f_{0001}\bigr).
	\end{aligned}\label{eq:mafcnt}
\end{gather}
\end{claim}

Theorem~\ref{tm:nullspace} below characterizes the set of  
identity coefficients that lead to the same distribution over joint biallelic genotypes. 
\begin{theorem}\label{tm:nullspace}
Suppose that $p+q=1$. 
If $\Delta_i\colon i=1,\dotsc,9$ satisfy~\eqref{eq:linear.twoalleles} then so do 
the following coefficients, for all choices of $\xi,\eta\in\mathbb{R}$. 
\begin{equation}\label{eq:nullspace}
\begin{aligned}
\Delta_1' & = \Delta_1-\eta pq &
\Delta_2' & = \Delta_2+\xi-\eta pq\\
\Delta_3' & = \Delta_3+2\eta pq &
\Delta_4' & = \Delta_4-\xi\\
\Delta_5' & = \Delta_5+2\eta pq &
\Delta_6' & = \Delta_6-\xi\\
\Delta_7' & = \Delta_7-\xi +\eta (1-2pq) \\
\Delta_8' & = \Delta_8+2\xi-2\eta &
\Delta_9' & = \Delta_9+\eta.
\end{aligned}
\end{equation}
Starting from an arbitrary particular 
solution set~$\Delta_i$, 
Equation~\eqref{eq:nullspace}
generates all vector solutions to~\eqref{eq:linear.twoalleles}. 
\end{theorem}

Values of~$(\xi,\eta)$ for which Eq.~\eqref{eq:nullspace} produces a proper distribution are 
precisely those where $\Delta_i'\ge 0$ for all~$i$: 
\begin{equation}\label{eq:nullspace.const}
\begin{aligned}
\eta & \le \frac{\Delta_1}{pq} & 
\eta & \le \frac{\Delta_2}{pq} + \frac{\xi}{pq}  &
\eta & \le \frac{\Delta_8}{2} + \xi \\ 
\eta & \ge -\frac{\Delta_3}{2pq} &
\eta & \ge -\frac{\Delta_5}{2pq} &
\eta & \ge -\frac{\Delta_7}{1-2pq} + \frac{\xi}{1-2pq} & 
\eta & \ge -\Delta_9 \\
\xi & \le \Delta_4 & 
\xi & \le \Delta_6 &
\end{aligned}
\end{equation}

\begin{figure}
\centerline{\includegraphics[width=\textwidth]{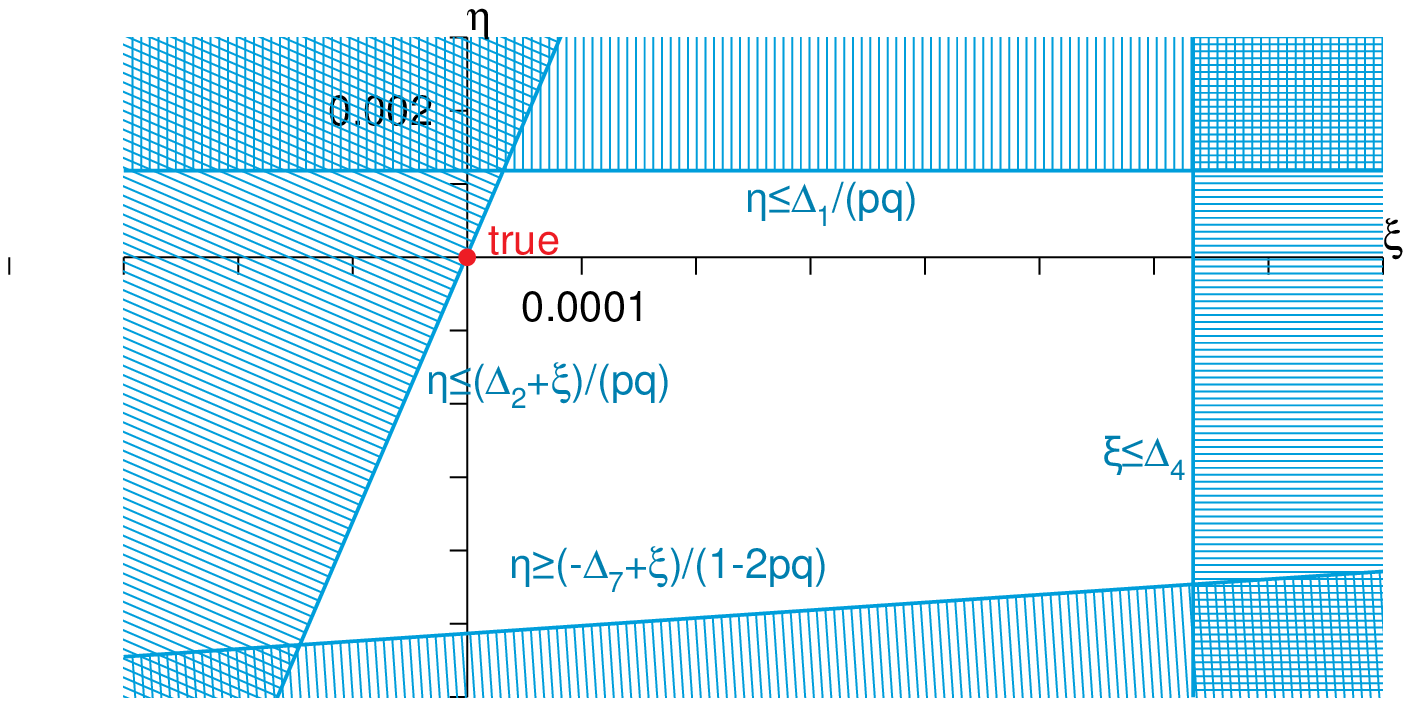}}
\begin{center}\footnotesize
\begin{tabular}{crccc}
\textbf{Identity coefficient} & $\times 2^{-20}$ 
	& \textbf{decimal value} & \multicolumn{2}{c}{\textbf{ambiguity range}}\\
	& & & min & max \\
$\Delta_1$ & 34 %& -14.9 
	& $3.24\cdot 10^{-5}$
	& $0$&$1.78\cdot 10^{-4}$\\
$\Delta_2$ & 1 %& -20 
	& $9.54\cdot 10^{-7}$
	& $0$&$7.58\cdot 10^{-4}$\\
$\Delta_3$ & 324 %& -11.7 
	& $3.09\cdot 10^{-4}$
	& $1.85\cdot 10^{-5}$&$3.74\cdot10^{-4}$\\
$\Delta_4$ & 665 %& -10.6 
	& $6.34\cdot 10^{-4}$
	& $0$&$7.80\cdot10^{-4}$\\
$\Delta_5$ & 3140 %& -8.4 
	& $2.99\cdot 10^{-3}$
	& $2.70\cdot 10^{-3}$&$3.06\cdot10^{-3}$\\
$\Delta_6$ & 9113 %& -6.8 
	& $8.69\cdot 10^{-3}$
	& $8.06\cdot 10^{-3}$&$8.84\cdot10^{-4}$\\
$\Delta_7$ & 5087 %& -7.7 
	& $4.85\cdot 10^{-3}$
	& $0$&$5.94\cdot10^{-3}$\\
$\Delta_8$ & 278698 %& -1.9 
	& $0.266$
	& $0.263$&$0.276$\\
$\Delta_9$ & 751514 %& 
	& $0.717$
	& $0.711$&$0.718$\\
\hline
\textbf{Relatedness parameter} \\
$\theta_1$ (kinship) & 73984 & $0.071$ \\
$\theta_{2A}$ (Victoria's inbreeding) & 1024 & $9.77\cdot 10^{-3}$\\
$\theta_{2B}$ (Albert's inbreeding) & 12288 & $0.0117$ \\
$\theta_{3}$ (two-out-of-three IBD) & 152824 & $0.146$ \\
$\theta_{3:3}$  (three-out-of-three IBD) & 1766 & $1.68\cdot 10^{-3}$ \\
$\theta_4$  $=(\Delta_4-\Delta_6)/2$ & -8448 & $-8.06\cdot 10^{-3}$\\
\hline
\end{tabular}
\end{center}

\caption[Null space identity coefficients]{Null space identity coefficients. 
	The intersection of the constraints from Equation~\eqref{eq:nullspace.const} 
	defines the convex polygonal area within which~$(\xi,\eta)$ values 
	plugged into Eq.~\eqref{eq:nullspace} 
	yield valid identity mode distributions that generate the same genotypic distribution.
	The example is based on the joint parentage of Queen Victoria of the United Kingdom (1819--1901) and 
	her spouse Prince Albert of Saxe-Coburg and Gotha (1819--1861), 
	for which the relevant parameters are listed below the plot.  
	The illustration assumes $pq=0.02746\dotsc$ 
	reflecting typical allele frequency moments 
	in humans
	(from the 1000 Genomes project). 
	Extremal values of possible~$\Delta_i$ 
	listed under ``ambiguity range'' are 
	attained in the corners of the unshaded area. 
}\label{fig:nullspace}
\end{figure}
Figure~\ref{fig:nullspace} illustrates the solution area of~\eqref{eq:nullspace.const} 
for the identity coefficients of a real-life example (Queen Victoria and Prince Albert, 
who shared multiple common ancestors within seven generations --- Appendix for complete 
family tree).

\subsection{Identifiable relatedness parameters}\label{ss:parameters}
Despite multiple solutions, some aspects of the identity coefficients can be ascertained from 
the genotype distribution. In particular, if 
a linear combination stays the same for all sets of identity coefficients  
from~\eqref{eq:nullspace}, then it is computable from the biallelic genotype distribution.   
Theorem~\ref{tm:parameter.identifiable} formalizes our argument. 

\begin{definition}
A function of the identity distribution~$\theta(\boldsymbol{\Delta})$ 
is called a {\em linear relatedness parameter} if and only if it can  be written as a linear combination 
\begin{equation}\label{eq:parameter.identifiable.def}
\theta(\boldsymbol{\Delta}) = \sum_{i=1}^9 a_i \Delta_i,  
\end{equation}
where~$a_i$ are constants. In particular, $a_i$ may not depend on the allele frequency~$p$. 
\end{definition}

\begin{theorem}\label{tm:parameter.identifiable}
A linear relatedness parameter~$\theta$ is identifiable from 
the joint genotype distribution only if 
\begin{equation}\label{eq:parameter.identifiable.condition}
\begin{aligned}
a_2 + 2a_8 & = a_4+a_6+a_7; \\
a_7+a_9 & = 2a_8; & \text{and}\\
2a_3+2a_5 & = a_1+a_2+2a_7.
\end{aligned}
\end{equation}
\end{theorem}

\begin{theorem}\label{tm:parameter.identifiable.list}
The following linear relatedness parameters are identifiable from the biallelic genotype distribution. 
\begin{subequations}\label{eq:parameter.identifiable.list}
\begin{align}
\theta_0 
	& = \sum_{i=1}^9 \Delta_i & \text{(=1)}\\
\theta_1
	& = \Delta_1 + \frac12\bigl(\Delta_3+\Delta_5+\Delta_7\bigr)+\frac14 \Delta_8 & \text{(kinship coefficient)}\\
\theta_{2\offather}
	& = \Delta_1+\Delta_2+\Delta_3+\Delta_4
		&\text{(A's inbreeding)}\\
\theta_{2\ofmother}
	& = \Delta_1+\Delta_2+\Delta_5+\Delta_6
		&\text{(B's inbreeding)}\\
\theta_3
	& = \begin{aligned}[t] 
		& \Delta_1+\Delta_2+\Delta_3+\Delta_5+\Delta_7\\
		+& \frac12\bigl(\Delta_4+\Delta_6+\Delta_8)
		\end{aligned}\\
\theta_4 
	& = \frac12\bigl(\Delta_4-\Delta_6	\bigr)
\end{align}
\end{subequations}
All other identifiable parameters are linear combinations of~$\theta_i$ in Equations~\eqref{eq:parameter.identifiable.list}. 
\end{theorem}

Theorem~\ref{tm:parameter.identifiable} shows that, in general, 
the identity coefficients~$\Delta_i$ are not identifiable separately.  
In particular, probabilities for various inbred modes ($\Delta_3,\Delta_4,\Delta_5,\Delta_6$) 
are not identifiable, only their differences ($\Delta_4-\Delta_6=2\theta_4$ and  $\Delta_3-\Delta_5=\theta_{2\offather}-\theta_{2\ofmother}-2\theta_4$).

The identifiable parameters of Theorem~\ref{tm:parameter.identifiable.list} include the usual measures of
inbreeding ($\theta_{2*}$) and coancestry ($\theta_1$) 
generalized to inbred parents~\cite{Harris.inbred.relatives}, as well 
as the trivial $\sum_i \Delta_i$. The parameter~$\theta_3$ is the probability that 
there is at least one pair of IBD alleles among three randomly selected ones. 
A simpler three-gene characterization of inbred coancestry is 
\begin{equation}\label{eq:trip3}
\theta_{3:3} = \Delta_1 + \frac12\bigl(\Delta_3+\Delta_5) = \theta_1 - \frac12 \theta_3 +\frac14\bigl(\theta_{2\offather}+\theta_{2\ofmother}\bigr),
\end{equation}
which is the probability that three randomly chosen alleles are simultaneously identical 
by descent. By Theorem~\ref{tm:parameter.identifiable}, $\theta_{3:3}$ is identifiable, and~\eqref{eq:trip3} shows 
how to write it as a linear combination of identifiable parameters from Theorem~\ref{tm:parameter.identifiable.list}.

Linear relatedness parameters, defined as linear combinations of identity coefficients, can 
be written
as linear combinations of genotypic probabilities in the 
linear-algebraic framework of Equation~\eqref{eq:jointcalls.p}.   
For the archetypical parameters of Theorem~\ref{tm:parameter.identifiable.list}, we consider the following 
expressions.
\begin{equation}\label{eq:statistics.def}
\begin{aligned}
\smpstat_{1\offather} & = \frac{f_{1101}+f_{0100}}2+ \frac{f_{0101}}4+ f_{1100} & 
\smpstat_{1\ofmother} & = \frac{f_{0111}+f_{0001}}2+ \frac{f_{0101}}4+ f_{0011}\\
\smpstat_1 & = \frac{\smpstat_{1\offather}+\smpstat_{1\ofmother}}2\\
\smpstat_{2\offather} & = \frac{f_{0111}+f_{0101}+f_{0100}}2 &
\smpstat_{2\ofmother} & = \frac{f_{1101}+f_{0101}+f_{0001}}2\\
\smpstat_{3} & = \frac{\bigl(f_{0100}-f_{0111}\bigr)+\bigl(f_{0001}-f_{1101}\bigr)}{4}
	& 
\smpstat_{4} & = \frac{f_{1100}-f_{0011}}{2}
\end{aligned}
\end{equation}
So, 
\begin{equation}\label{eq:statistics.to.parameters}
\begin{aligned}
\theta_1 & = 1-\frac{\smpstat_{1\offather}}{p-p^2} = 1-\frac{\smpstat_{1\ofmother}}{p-p^2} =1-\frac{\smpstat_{1}}{p-p^2}\\
\theta_{2\offather} & = 1-\frac{\smpstat_{2\offather}}{p-p^2} & \theta_{2\ofmother} & = 1-\frac{\smpstat_{2\ofmother}}{p-p^2}\\
\theta_3 & = 1-\frac{\smpstat_{3}}{p-3p^2+2p^3} & 
\theta_4 & = \frac{\smpstat_4}{p-3p^2+2p^3}.
\end{aligned}
\end{equation}
Due to the linear dependencies, multiple equivalent formulas exist 
that relate the genotype distribution and any specific parameter. 
For instance, the intermediate quantities $\tau_{1\offather}$, $\tau_{1\ofmother}$ and~$\tau_1$, which 
weigh genotypic probabilities differently, 
are equal by Equation~\eqref{eq:mutantdiff}.

\subsection{Relatedness estimation from independent loci} \label{ss:estimation}
We examine two applications of estimating pairwise relatedness from 
observed genotypes at independent diallelic loci. 
\paragraph{Application~I: relatedness between two individuals} 
In this application~\cite{Ritland.relatedness.MME,LynchRitland.relatedness}, 
$n$ independent sites are genotyped in two genomes. Minor-allele frequencies~$p_i$  
follow population-wide background frequencies, and apply to both genomes 
equally. Our aim is to characterize the IBD mode distribution implied by the joint genealogy,
using the observed genotypes. 

\paragraph{Application~II: background relatedness in a structured population} 
We sample a set of~$n$ random pairs from a population at the same locus, and we 
would like to infer the background structure of relatedness~\cite{AndersonWeir.relatedness.structured.ML,Wang.structured.population}, 
described as the population-wide distribution of IBD modes.
Accordingly, the same minor-allele frequency $p_i\equiv p$ applies to all pairs~$i=1,2,\dotsc,n$.   

Theorem~\ref{tm:nullspace} suggests that,
identity coefficients cannot always be inferred 
from observed genotypes in either application.  
We formalize our argument using the following abstraction. 
Suppose that minor-allele frequencies~$p_i\colon i=1,\dotsc,n$ apply 
to~$n$ independent biallelic loci sampled by random  
pairwise genotypes~$X_i\in\{\mathsf{0/0}, \mathsf{0/1}, \mathsf{1/1}\}^2\colon i=1,2,\dotsc,n$. 
The pairs have identical 
IBD mode distribution~$\boldsymbol{\Delta}$. 
As in Eq.~\eqref{eq:linear.twoalleles}, 
the joint genotype distribution~$\mathbf{f}^{(i)}$ of~$X_i$ is related to the 
identity coefficients by a matrix~$\mfreq^{(i)}$ through $\mathbf{f}^{(i)} = \mfreq^{(i)} \cdot \boldsymbol\Delta$.
The matrix entries are given by~\eqref{eq:jointcalls.p}, substituting~$p\leftarrow p_i$ and
$q\leftarrow 1-p_i$. 

\begin{theorem}\label{tm:notconsistent}
Let~$T_n(X_1,X_2,\dotsc, X_n)$ be an estimator of the identity coefficients~$\boldsymbol{\Delta}$. 
If~$\min\{\Delta_4,\Delta_6,\Delta_7\}>0$ 
or $\min\{\Delta_2,\Delta_8\}>0$
or both, then~$T_n$ may not converge (in probability) to~$\boldsymbol{\Delta}$ as~$n\to\infty$
for any sequence of minor-allele frequencies~$p_i\colon i=1,2,\dotsc$. 
\end{theorem}

In other words, identity coefficients cannot be estimated consistently in general. 
Theorem~\ref{tm:parameter.identifiable.list} suggests, however, that 
a maximal set of linear relatedness parameters can be inferred consistently.
As an illustration, we prove the consistency of simple moment-based estimators
using observed genotype frequencies. 
Define the genotype counts~$n_{0000},n_{1111},n_{1101},\dotsc,n_{0001}$, 
and the {\em empirical genotype frequencies} $\hat{f}_x = n_x/n$. Written 
with indicator variables $\{X_i=x\}$ for $x=0000, 1111, \dotsc, 0001$, 
\begin{equation}\label{eq:f.sum.indicator}
\hat{f}_x = \frac{n_x}n = \frac{\sum_{i=1}^n \bigl\{ X_i=x\bigr\}}n. 
\end{equation}

Plugged into Equation~\eqref{eq:statistics.def}, we get the sample statistics 
\begin{equation}\label{eq:statistics.empirical}
\begin{aligned}
\hat{\smpstat}_{1\offather} & = \frac{\hat{f}_{1101}+\hat{f}_{0100}}2+ \frac{\hat{f}_{0101}}4+ \hat{f}_{1100} \\ 
\hat{\smpstat}_{1\ofmother} & = \frac{\hat{f}_{0111}+\hat{f}_{0001}}2+ \frac{\hat{f}_{0101}}4+ \hat{f}_{0011} &
\hat{\smpstat}_1 & = \frac{\hat{\smpstat}_{1\offather}+\hat{\smpstat}_{1\ofmother}}2\\
\hat{\smpstat}_{2\offather} & = \frac{\hat{f}_{0111}+\hat{f}_{0101}+\hat{f}_{0100}}2 &
\hat{\smpstat}_{2\ofmother} & = \frac{\hat{f}_{1101}+\hat{f}_{0101}+\hat{f}_{0001}}2\\
\hat{\smpstat}_{3} & = \frac{\bigl(\hat{f}_{0100}-\hat{f}_{0111}\bigr)+\bigl(\hat{f}_{0001}-\hat{f}_{1101}\bigr)}{4}
	& 
\hat{\smpstat}_{4} & = \frac{\hat{f}_{1100}-\hat{f}_{0011}}{2}.
\end{aligned}
\end{equation}
Write the moments for the minor-allele frequency (MAF) distribution as
\[
\begin{aligned}
\npsnp & = \frac1n \sum_{i=1}^{n} p_i & \nqsnp & = \frac1n \sum_{i=1}^n p_i^2 &
\ntsnp & = \frac1n \sum_{i=1}^{n} p_i^3 & \nfsnp & = \frac1n \sum_{i=1}^n p_i^4,
\end{aligned}
\]
and the average centralized moments as
\[
\begin{aligned}
\ndeux & = \frac1n \sum_{i=1}^n \Bigl( p_i(1-p_i)^2 + (1-p_i)p_i^2 \Bigr) = \npsnp-\qsnp\\
\ntrois & = \frac1n \sum_{i=1}^n \Bigl( p_i(1-p_i)^3 - (1-p_i)p_i^3 \Bigr)= \npsnp-3\nqsnp+2\ntsnp.
\end{aligned}
\]
Finally, define the relatedness estimators
\begin{equation}\label{eq:estimator.def}
\begin{aligned}
\hat{\theta}_1 & = 1-\frac{\hat{\smpstat}_{1}}{\ndeux} &
\hat{\theta}_{2\offather} & = 1-\frac{\hat{\smpstat}_{2\offather}}{\ndeux} & \hat{\theta}_{2\ofmother} & = 1-\frac{\hat{\smpstat}_{2\ofmother}}{\ndeux}\\
\hat{\theta}_3 & = 1-\frac{\hat{\smpstat}_{3}}{\ntrois} & 
\hat{\theta}_4 & =\frac{\hat{\smpstat}_{4}}{\ntrois}.
\end{aligned}
\end{equation}
\begin{theorem}\label{tm:consistent.parameters}
Suppose that 
there exist constants $\cdeux,\ctrois$ 
such that 
in Eq.~\eqref{eq:estimator.def},
	$\ndeux\convprob\cdeux$ (converges in probability to $\cdeux$)
	and
	$\ntrois\convprob\ctrois$
	as $n\to\infty$.
	If $\cdeux,\ctrois\ne 0$, 
	then $\hat{\theta}_y\convprob \theta_y$ 
	for all parameters $y=1,2\offather,2\ofmother,3,4$ from Theorem~\ref{tm:parameter.identifiable.list}. 
\end{theorem}
The theorem implies that one can consistently estimate five non-trivial parameters 
of the background IBD structure in a population (Application II) 
if the sampled locus has minor-allele frequency $p\ne 0,1,1/2$. 
In Application~I, our moment-based estimators are consistent when, for instance,
minor-allele frequencies are independent and identically distributed, with 
non-zero variance (i.e., sites are mostly segregating) 
and kurtosis (excluding symmetric MAF distributions around 1/2). 

\section{Discussion}
Identity coefficients~\cite{Jacquard,Harris.inbred.relatives} encapsulate 
the allele dependencies within the joint genotype distribution for a pair of 
diploid individuals 
If the individuals are not inbred, only three of the 
coefficients may be positive ($\Delta_7,\Delta_8,\Delta_9$), corresponding 
to Cotterman's $k$-gene coefficients~\cite{Thompson.pairwise}
for the individuals sharing~$k=0$, $1$, or~$2$ alleles between them.  
The three coefficients can be retrieved
from sampled genotypes 
using well-established methods 
relying on 
likelihood maximization~\cite{Thompson.pairwise,Milligan.relatedness.ML} 
or allele frequency moments~\cite{Ritland.relatedness.MME,LynchRitland.relatedness}. 

It may be of interest to estimate higher-order identity coefficients 
simultaneously. In particular, all nine IBD modes may occur if the individuals 
have inbred coancestries~\cite{Harris.inbred.relatives}, or come from 
a structured population~\cite{Wang.structured.population}. 
Biallelic genotypes, however, do not convey enough information 
about the generic IBD structure, since different 
identity coefficients can generate the same joint genotype distribution. 
Theorem~\ref{tm:nullspace} scrutinizes the inherent ambiguity 
about the identity coefficients, describing the linear subspace in which 
all solutions are found. 
One particular source of the ambiguity 
(corresponding to the null vector~$\mathbf{z}_1$ in~\eqref{eq:nullspace.vectors}) 
is that symmetric mixtures of simultaneous 
inbreeding and coancestry (modes $\Delta_7$-$\Delta_4$-$\Delta_6$ vs.\ 
$\Delta_2$-$\Delta_8$-$\Delta_8$ in Figure~\ref{fig:coancestry}) manifest 
identically in the genotype distribution. 
Importantly, these equivalent solutions (varying only 
the~$\xi$ coordinate) 
remain equivalent for any minor-allele frequency.
The uncertainties about the identity coefficients 
are within the same magnitude as the inbreeding levels 
(Eq.~\eqref{eq:nullspace})
when both individuals are inbred ($\Delta_4,\Delta_6> 0$)
and share multiple ancestors ($\Delta_7>0$). 
Indeed, the real-life example of Figure~\ref{fig:nullspace} 
shows that the subtle details of coancestry can be irretrievable 
from the genotype distribution.

Consistent estimation is thus impossible 
since even as the number of independent sampled loci~$n$ goes to infinity
and genotype frequencies concentrate around their true probabilities, 
the identity coefficients stay ambiguous regardless of the estimation method used (Theorem~\ref{tm:notconsistent}). 
The decomposition of the solution space (Theorem~\ref{tm:parameter.identifiable}) shows the 
aspects of the IBD structure that can instead be inferred from biallelic genotypes. 
Specifically, Theorem~\ref{tm:parameter.identifiable.list} lists five non-trivial 
relatedness parameters, deconvolving the IBD structure to the maximum degree that is attainable. 
Principal aspects of genetic relatedness, quantified by the coefficients of kinship and inbreeding, are 
identifiable. Other identifiable attributes are probabilities for three-gene joint IBD and the 
asymmetry of inbreeding modes with and without simultaneous coancestry. In contrast, 
parameters that do not weigh the identity coefficients properly (Eq.~\eqref{eq:parameter.identifiable.condition})
are not identifiable from the biallelic genotypes. Ill-defined relatedness parameters include
the probabilities of separate identity modes (e.g., the probability~$\Delta_1$ of fourfold IBD),
the fraternity coefficient ($\Delta_1+\Delta_7$) and other generalizations 
of Cotterman's $k$-gene coefficients.

In the case of human genomes, the limits of inferring relatedness 
are set by linkage and 
finite genome size, and not identifiability~\cite{Skare.distant}.  
The mean length of a segment with the same particular history involving~$m$ meioses 
decreases linearly with~$m$. 
In human whole genome sequences, IBD segments of length~$0.4\,\mathrm{cM}$ 
can be demarcated~\cite{Su.IBD.segmentlength} with confidence by 
high-coverage sequencing. 
The detection of shared ancestry is thus constrained by the fact that  
descendants inherit a common ancestor's allele simultaneously with exponentially 
small probability in the number of meioses separating them ($2^{-m+1}$).
As Browning and Browning~\cite{Browning.ibd.review} point out, 
fifth cousins ($m=12$) simultaneously inherit~$1/2048$ 
of their genome on expectation from the shared
great-great-great-great grandfather or great-great-great-great grandmother 
each, which amounts to about~$1.5\,\mathrm{cM}$
in an entire human genome, while the average 
IBD segment length is $8.3\,\mathrm{cM}$. 
Then, by Markov's inequality, there is at least one IBD segment between the 
two cousins' genomes with probability at most~$\frac{2\times 1.5}{8.3}=0.35\dotsc$. 
Identity modes with more than two IBD alleles 
($\Delta_1,\Delta_2,\Delta_3,\Delta_5,\Delta_7$)
usually involve even more distant ancestries and are, 
thus, almost certainly undetectable~\cite{Thompson.process}. 
For example, if both individuals are children of fourth cousins
(as the royal cousins here), 
the simultaneous inbreeding mode~$\Delta_2$ 
appears in segments of average length~$4.2\,\mathrm{cM}$, 
covering $1/2^{20}$ of their genome (about $0.003\,\mathrm{cM}$); so, 
no such segments are seen in at least~$99.93\%$ of the cases.

The ambiguity of identity coefficients (outlined by Theorem~\ref{tm:nullspace}) 
complements well-known results on equivalent pedigrees~\cite{Donnelly,Skare.distant}.
In the absence of linkage information, 
the non-identifiable mode combinations represent the theoretical limits of dissecting 
the IBD structure. By our results (Theorem~\ref{tm:parameter.identifiable.list}), 
only two more distribution parameters can be inferred
in addition to the usual two-gene coefficients for coancestry and inbreeding: one for three-gene 
IBD, and another measuring asymmetry in inbreeding modes.

\section*{Acknowledgments}
I am grateful for valuable commentary on earlier versions of the manuscript by Damian Labuda and 
anonymous reviewers. This research project was partially funded by the author's individual discovery grant 
from the 
{\em Natural Sciences and Engineering Research Council of Canada}.

\clearpage
\appendix
\section{Linear algebra for genotypic probabilities and identity coefficients}

\begin{proof}[ of Claim~\ref{cm:mafcnt}]
The equalities can be seen by inspecting the rows of~$\mfreq$, but considering allele counts gives a more straightforward proof. 
Consider the expected number~$\mafcnt$ of minor ('1') alleles in the random joint genotype. 
Since it is the expectation for the sum of four indicator variables, $\EXP\mafcnt = 4 p$.   
Alternatively, by summing over the possible joint genotypes, 
$\EXP\mafcnt = 4f_{1111} + 3 \bigl(f_{1101}+f_{0111}\bigr)+2 \bigl(f_{0101}+f_{1100}+f_{0011}\bigr)+\bigl(f_{0100}+f_{0001}\bigr)$, 
and Equation~\eqref{eq:mafcnt} follows after dividing by~4. 
Now consider the expected number of minor alleles in the 
A's and B's genotype separately. Clearly, both equal~$2p$. Counting by joint genotypes: 
\begin{multline*}
\underbrace{2\bigl(f_{1111}+f_{1101}+f_{1100}\bigr) + \bigl(f_{0111}+f_{0101}+f_{0100}\bigr)}_{\text{expected count in A}} \\
=  \underbrace{2\bigl(f_{1111}+f_{0111}+f_{0011}\bigr) + \bigl(f_{1101}+f_{0101}+f_{0001}\bigr)}_{\text{expected count in B}}.
\end{multline*}
After elimination of common terms, Equation~\eqref{eq:mutantdiff} follows. 
\end{proof}

\begin{proof}[ of Theorem~\ref{tm:nullspace}]
The null space of~$\mfreq$ is spanned by the vectors
\begin{align} \label{eq:nullspace.vectors}
\mathbf{z}_1 & 
	= \begin{pmatrix}
		0\\
		1\\
		0\\
		-1\\
		0\\
		-1\\
		-1\\
		2\\
		0
	\end{pmatrix} 
&
\mathbf{z}_2 & 
= \begin{pmatrix}
	0\\
	0\\
	0\\
	0\\
	0\\
	0\\
	1\\
	-2\\
	1
	\end{pmatrix}
	+ pq 
	\begin{pmatrix}
	-1\\
	-1\\
	2\\
	0\\
	2\\
	0\\
	-2\\
	0\\
	0
	\end{pmatrix}
	= \mathbf{z}_{2}^{(1)}+pq \mathbf{z}_{2}^{(2)}
\end{align}
It is straightforward to verify that $\mfreq\mathbf{z}_1=\mfreq\mathbf{z}_2=\mathbf{0}$, the null vector.  
Hence, 
\[
\mfreq\cdot \bigl(\mathbf{\Delta}+\xi\mathbf{z}_1+\eta\mathbf{z}_2\bigr) = \mfreq\cdot \mathbf{\Delta}=\mathbf{f}
\]
for all choices of $\xi,\eta$. The rank of the~$9\times9$ 
matrix~$\mfreq$ is~7 (established by Gaussian elimination), and therefore no other solutions exist. 
\end{proof}

\begin{proof}[ of Theorem~\ref{tm:parameter.identifiable}]
In order to be identifiable, $\theta(\boldsymbol{\Delta})$ must remain the same 
for all distributions satisfying~\eqref{eq:linear.twoalleles}. 
The vector of coefficients~$(a_i\colon i=1,\dotsc,9)$ then has to be orthogonal 
to the null space of~$\mfreq$. The identities of~\eqref{eq:parameter.identifiable.condition} express 
the orthogonality with the vectors~$\mathbf{z}_1$, $\mathbf{z}_2^{(1)}$ and~$\mathbf{z}_2^{(2)}$:
the latter two are used separately since orthogonality must be maintained for all~$p$. 
\end{proof}

\begin{proof}[ of Theorem~\ref{tm:parameter.identifiable.list}]
By Theorem~\ref{tm:parameter.identifiable}, 
identifiable parameters satisfy three independent linear equations. 
The theorem lists a maximal set of~6 linearly independent parameters.
\end{proof}

\section{Genotype counts and MAF moments}

\begin{proof}[ of Theorem~\ref{tm:notconsistent}]
By Eq.~\eqref{eq:nullspace.const}, a proper distribution~$\boldsymbol{\Delta}'$
is generated in Eq.~\eqref{eq:nullspace} 
 with~$\eta=0$ and all~$\xi$ satisfying
\[
-\min\{\Delta_2,\Delta_8/2\}\le \xi \le \min\{\Delta_4,\Delta_6,\Delta_7\}.
\]
If~$\xi\ne 0$, then $\boldsymbol{\Delta}'\ne\boldsymbol{\Delta}$, yet they 
generate the same joint genotype distributions $\mathbf{f}^{(i)}$ 
with all minor-allele frequency sequences~$(p_i\colon i=1,2,\dotsc)$.
\end{proof}

\begin{proof}[ of Theorem~\ref{tm:consistent.parameters}]
Define the expectations $\avgfreq_x = \EXP \hat{f}_x$. 
Since~$\hat{f}_x$ is 
an average of indicator random variables in~\eqref{eq:f.sum.indicator},
\begin{equation}\label{eq:f.convprob}
\hat{f}_x\convprob \bar{f}_x
\end{equation}
as~$n\to\infty$. 
(Hoeffding's inequality gives $\PROB\Bigl\{ \bigl|\hat{f}_x-\bar{f}_x\bigr| \ge \epsilon\Bigr\} \le 2\exp(-2n\epsilon^2)$, thus 
$\lim_{n\to\infty} \PROB\Bigl\{ \bigl|\hat{f}_x-\bar{f}_x\bigr| \ge \epsilon\Bigr\} = 0$ for all~$\epsilon>0$.)

By averaging~\eqref{eq:jointcalls.p} across the~$n$ sites, 
\[
\bar{\mathbf{f}} = \frac{\sum_{i=1}^n \mathbf{f}^{(i)}}{n} = \frac{\sum_{i=1}^n \mfreq^{(i)}}{n}\cdot\boldsymbol{\Delta} = \bar{\mfreq} \cdot \boldsymbol\Delta.
\]
The entries of~$\bar{\mfreq}$ link 
expected genotype frequencies~$\avgfreq$ to the common identity coefficients~$\Delta_i\colon i=1,\dotsc,9$, 
as follows in Eqs.~(\ref{eq:jointcalls.moments.0000}--\ref{eq:jointcalls.moments.0001}).

%\begin{equation}
\begin{subequations}\label{eq:jointcalls.moments}
\begin{align}
\avgfreq_{0000} & = \frac{\EXP n_{0000}}{n}
	= \begin{aligned}[t]
	& (1-\npsnp)\Delta_1 
	+ (1-2\npsnp+\nqsnp)\bigl(\Delta_2+\Delta_3+\Delta_5 +\Delta_7\bigr)\\
	+ & (1-3\npsnp+3\nqsnp-\ntsnp)\bigl(\Delta_8+\Delta_4+\Delta_6\bigr)\\
	+& (1-4\npsnp+6\nqsnp-4\ntsnp+\nfsnp)\Delta_9
	\end{aligned}\label{eq:jointcalls.moments.0000}\\
\avgfreq_{1111} & = \frac{\EXP n_{1111}}{n}
	= \begin{aligned}[t]
	& \npsnp\Delta_1 + \nqsnp\bigl(\Delta_2+\Delta_3+\Delta_5 +\Delta_7\bigr)\\
	+& \ntsnp\bigl(\Delta_4+\Delta_6+\Delta_8\bigr)+\nfsnp\Delta_9
	\end{aligned} \\
\avgfreq_{1101} & = \frac{\EXP n_{1101}}{n}
	= \begin{aligned}[t] 
		& (\npsnp-\nqsnp)\Delta_3+(\nqsnp-\ntsnp)(\Delta_8+2\Delta_4)\\
		+& 2(\ntsnp-\nfsnp)\Delta_9 
		\end{aligned}\\
\avgfreq_{0111} & = \frac{\EXP n_{0111}}{n}
	= \begin{aligned}[t] 
		& (\npsnp-\nqsnp)\Delta_5 +(\nqsnp-\ntsnp)(\Delta_8+2\Delta_6)\\
		+& 2(\ntsnp-\nfsnp)\Delta_9 
		\end{aligned}\\
\avgfreq_{0101} & = \frac{\EXP n_{0101}}{n}
	= \begin{aligned}[t]
		& 2(\npsnp-\nqsnp)\Delta_7+ (\npsnp-2\nqsnp+\ntsnp)\Delta_8\\
		+& 4(\nqsnp-2\ntsnp+\nfsnp)\Delta_9
		\end{aligned}\\
\avgfreq_{1100} & = \frac{\EXP n_{1100}}{n}
	= \begin{aligned}[t] 
	& (\npsnp-\nqsnp)\Delta_2 + (\npsnp-2\nqsnp+\ntsnp)\Delta_4+(\nqsnp-\ntsnp)\Delta_6\\
	+& (\nqsnp-2\ntsnp+\nfsnp)\Delta_9 
	\end{aligned}\\
\avgfreq_{0011} & = \frac{\EXP n_{0011}}{n}
	= \begin{aligned}[t] 
	& (\npsnp-\nqsnp)\Delta_2 + (\npsnp-2\nqsnp+\ntsnp)\Delta_6+(\nqsnp-\ntsnp)\Delta_4\\
	+& (\nqsnp-2\ntsnp+\nfsnp)\Delta_9 
	\end{aligned}\\
\avgfreq_{0100} & = \frac{\EXP n_{0100}}{n}
	= \begin{aligned}
	& (\npsnp-\nqsnp)\Delta_5 + (\npsnp-2\nqsnp+\ntsnp)(\Delta_8 + 2\Delta_6) \\
	+&  2(\npsnp-3\nqsnp+3\ntsnp-\nfsnp)\Delta_9
	\end{aligned}\\
\avgfreq_{0001} & = \frac{\EXP n_{0001}}{n} 
	= \begin{aligned}[t] 
		& (\npsnp-\nqsnp)\Delta_3 + (\npsnp-2\nqsnp+\ntsnp)(\Delta_8 + 2\Delta_4) \\
		+& 2(\npsnp-3\nqsnp+3\ntsnp-\nfsnp)\Delta_9.
		\end{aligned}\label{eq:jointcalls.moments.0001}
\end{align}
\end{subequations}

In a similar vein, define the expected sample statistics $\bar{\smpstat}_y=\EXP \hat{\smpstat}_y$ 
for $y=1,2\offather,2\ofmother,3,4$. It follows from~\eqref{eq:statistics.empirical} and~\eqref{eq:statistics.def}
that each~$\bar{\smpstat}_y$ can be written as a linear combination of the expected genotype counts~$\bar{f}_x$; i.e., 
Equation~\eqref{eq:statistics.def} holds after substituting $\smpstat_y \leftarrow \bar{\smpstat}_y$ and $f_x\leftarrow \bar{f}_x$
appropriately. Consequently, \eqref{eq:jointcalls.moments} and~\eqref{eq:f.convprob} imply that 
\begin{equation}\label{eq:estimator.convprob}
\begin{aligned}
\hat{\smpstat}_{1} & \convprob (\npsnp-\nqsnp) (1-\theta_1) \\ 
\hat{\smpstat}_{2\offather} & \convprob (\npsnp-\nqsnp) (1-\theta_{2\offather})
& \hat{\smpstat}_{2\ofmother} & \convprob (\npsnp-\nqsnp) (1-\theta_{2\ofmother})\\
\hat{\smpstat}_{3} & \convprob (\npsnp-3\nqsnp+2\ntsnp) (1-\theta_3) & 
\hat{\smpstat}_{4} & \convprob (\npsnp-3\nqsnp+2\ntsnp) \theta_4.
\end{aligned}
\end{equation}
Since 
\[
\npsnp-\nqsnp\convprob \cdeux
\qquad \text{and} \qquad 
(\npsnp-3\nqsnp+2\ntsnp)\convprob \ctrois,
\]
and the convergence in the denominators of~$\hat{\theta}_y$ is 
towards non-zero constants ($\cdeux$ or $\ctrois$), 
Eq.~\eqref{eq:estimator.convprob} implies $\hat{\theta}_y\convprob \theta_y$
for all $y=1,2\offather,2\ofmother,3,4$. 
\end{proof}

\section{Family tree of Queen Victoria and Prince Albert} 
\begin{figure}
\centerline{\includegraphics[width=\textwidth]{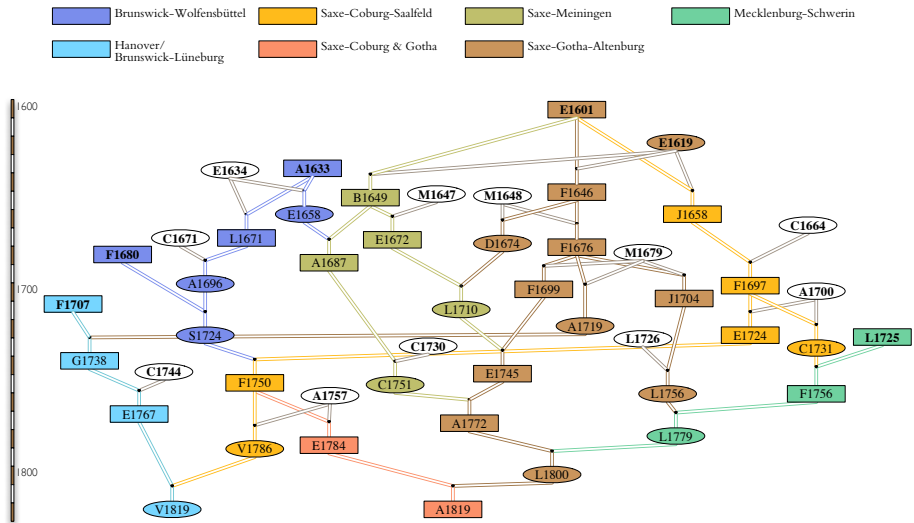}}

\caption[Family tree of Queen Victoria and Prince Albert]{Joint ancestry of Queen Victoria (V1819) and Prince Albert (A1819).
	Rectangles denote men; ovals denote women. 
	Nodes are placed by the indicated time scale, according to birth year. 
}\label{fig:tree}
\end{figure}

Queen Victoria of the United Kingdom (1819--1901) and her spouse Prince Albert of Saxe-Coburg and Gotha (1819--1861) 
descended from prominent ducal families of the Holy Roman Empire. Figure~\ref{fig:tree} 
shows their family tree covering 200 years, pruned back to founders. Table~\ref{tbl:ancestors} 
lists the 49 members of the tree. 
Genealogical records were culled using \url{peerage.com}, and Wikipedia's 
English and German editions. 

{\footnotesize
\begin{longtable}[t]{lccll}
\caption[Family members for Victoria and Albert's genealogy.]{Members of the family tree for Victoria and Albert. Identifiers 
	are formed by initial and birth year; founders are marked by boldface.}\label{tbl:ancestors}\\
\hline
\textbf{Identifier} & \multicolumn{2}{c}{\textbf{Lineage}} & \textbf{Name} & \textbf{Title} \\
	& {\footnotesize Victoria} & {\footnotesize Albert} 
\endfirsthead
\caption*{Table~\thetable. Family tree for Victoria and Albert}\\
\hline
\textbf{Identifier} & \multicolumn{2}{c}{\textbf{Lineage}} & \textbf{Name} & \textbf{Title} \\
	& {\footnotesize Victoria} & {\footnotesize Albert} 
\endhead

\textbf{E1601}	& \female\male & \female\male & Ernest I	& Duke of Saxe-Gotha and Altenburg	\\
\textbf{E1619}	& \female\male & \female\male 	& Elisabeth Sophie	& of Saxe-Altenburg\\
\textbf{A1633}	& \female & \female\male	& Anthony Ulrich	& Duke of Brunswick-WolfenbŸttel	\\
\textbf{E1634}	& \female & \female\male & Elisabeth Juliane	& of Schleswig-Holstein-S{\o}nderborg-Nordborg  \\
F1646		& \male  & \female & Frederick I	& Duke of Saxe-Gotha-Altenburg\\
\textbf{M1647}	& & \female& Marie Hedwig	& of Hesse-Darmstadt	\\
\textbf{M1648}	& \male & \female& Magdalena Sibylle	& of Saxe-Weissenfels 	\\
B1649	&	 & \female & Bernhard I	& Duke of Saxe-Meiningen \\
E1658	&  & \female & Elisabeth Eleonore	& of Brunswick-Wolfenb\"uttel	\\
J1658	& \female & \female\male& John Ernest IV	& Duke of Saxe-Coburg-Saalfeld	 \\
\textbf{C1664}	& \female & \female\male & Charlotte-Johanna	& of Waldeck-Wildungen \\
\textbf{C1671}	& \female 	& \male& Christine Louise	& of Oettingen-Oettingen \\
L1671	& \female & \male & Louis Rudolph	& Duke of Brunswick-L\"uneburg  \\
E1672	& & \female& Ernst Ludwig I	& Duke of Saxe-Meiningen \\
D1674	& & \female& Dorothea Marie	& of Saxe-Gotha-Altenburg	\\
F1676	& \male & \female & Frederick II	& Duke of Saxe-Gotha-Altenburg	\\
\textbf{M1679}	& \male & \female & Magdalene Augusta	&  of Anhalt-Zerbst	\\
\textbf{F1680}	& \female & \male& Ferdinand-Albert II	& Duke of Brunswick-Wolfenb\"uttel \\
A1687	 & & \female & Anton Ulrich	& Duke of Saxe-Meiningen\\
A1696	& \female & \male	& Antoinette	& of Brunswick-Wolfenb\"uttel \\
F1697	& \female & \female\male & Francis Josias & Duke of Saxe-Coburg-Saalfeld \\
F1699	& & \female  & Frederick III	& Duke of Saxe-Gotha-Altenburg	\\
\textbf{A1700}	& \female & \female\male& Anna Sophie	& of Schwarzburg-Rudolstadt	\\
J1704		& & \female & John August & of Saxe-Gotha-Altenburg\\
\textbf{F1707}	& \male & & Frederick	Louis & Prince of Wales	\\
L1710	& & \female & Luise Dorothea	& of Saxe-Meiningen	\\
A1719	& \male &  & Augusta	& of Saxe-Gotha-Altenburg	\\
E1724	 & \female & \male& Ernest Frederick	& Duke of Saxe-Coburg-Saalfeld	\\
S1724	& \female & \male & Sophie Antoinette	& of Brunswick-Wolfenb\"uttel 	\\
\textbf{L1725}	& & \female & Louis	& Duke of Mecklenburg-Schwerin	 \\
\textbf{L1726}	& & \female & Louise	& of Reuss-Schleiz  \\
\textbf{C1730}	 & & \female	& Charlotte Amalie 	& of Hesse-Phillipstal\\
C1731	& & \female & Charlotte Sophie	& of Saxe-Coburg-Saalfeld \\
G1738	& \male & & George III	& King of the United Kingdom \\
\textbf{C1744}	& \male && Charlotte	& of Mecklenburg-Strelizt	 \\
E1745	& & \female & Ernest II	& Duke of Saxe-Gotha-Altenburg	\\
F1750	& \female & \male& Francis 	& Duke of Saxe-Coburg-Saalfeld	 \\
C1751		& & \female & Charlotte	& of Saxe-Meiningen \\
F1756	 & & \female & Frederick Francis I	& Grand-Duke of Mecklenburg-Schwerin\\
L1756	& & \female & Louise	& of Saxe-Gotha-Altenburg	\\
\textbf{A1757}	& \female & \male & Augusta 	& of Reuss-Ebersdorf \\
E1767	& \male & & Edward	& Duke of Kent and Strathearn \\
A1772	& & \female & Augustus & Duke of Saxe-Gotha-Altenburg \\
L1779	 & & \female & Louise-Charlotte	& of Mecklenburg-Schwerin \\
E1784		& & \male & Ernest I	& Duke of Saxe-Coburg and Gotha\\
V1786	& \female & & Victoria & of Saxe-Coburg-Saalfeld \\
L1800	& & \female & Louise	& of Saxe-Gotha-Altenburg  \\
A1819	& & & Albert	& of Saxe-Coburg and Gotha\\
V1819	& & & Victoria	& Queen of the United Kingdom\\
\hline
\end{longtable}
}

\end{document}